\documentclass[epj]{svjour}
\usepackage{graphics}
\usepackage{amsmath}
\RequirePackage{graphicx}

\usepackage{hyperref}
\hypersetup{
    colorlinks=true,
    linkcolor=blue,
    filecolor=blue,
    urlcolor=blue,
}

\begin{document}
\title{New apparatus design for high precision measurement of G with atom interferometry}

\author{M. Jain\inst{1} \and G. M. Tino\inst{1} \thanks{Also at CNR-IFAC, Sesto Fiorentino, Italy.} \and L. Cacciapuoti\inst{2} \and G. Rosi\inst{1} \thanks{rosi@fi.infn.it}}                     
%
%
\institute{Dipartimento di Fisica e Astronomia and LENS, Universit\`{a} di Firenze, INFN Sezione di Firenze, via Sansone 1, I-50019 Sesto Fiorentino (FI), Italy \and European Space Agency, Keplerlaan 1, 2200 AG Noordwijk, The Netherlands}

\abstract{
We propose a new scheme for an improved determination of the Newtonian gravitational constant $G$ and evaluate it by numerical simulations. Cold atoms in free fall are probed by atom interferometry measurements to characterize the gravitational field generated by external source masses. Two source mass configurations having different geometry and using different materials are compared to identify an optimized experimental setup for the $G$ measurement. The effects of the magnetic fields used to manipulate the atoms and to control the interferometer phase are also characterized.
} 
\maketitle
\section{Introduction and Motivation}

Atom interferometry \cite{tino2014atom} has gained prominence due to a number of successful experiments carrying out precise measurements of the gravitational acceleration \cite{kasevich1992measurement,peters1999measurement,mueller2008atom,le2008limits,d2019measuring}, curvature \cite{rosi2015measurement,asenbaum2017phase}, gravity gradient \cite{mcguirk2002sensitive,snadden1998measurement,sorrentino2014sensitivity,duan2014operating,fpdphysrevA,wang2016extracting,d2017canceling}, rotations \cite{gustavson2000rotation,gustavson1997precision,canuel2006six,gauguet2009characterization}, gravitational red shifts \cite{hohensee2012force,muller2010precision,wolf2010atom} as well as fundamental constants like $h/m$ \cite{cadoret2009atom,clade2016precise,weiss1993precision,gupta2002contrast} and the Newtonian gravitational constant \cite{lamporesi2007source,lamporesi2008determination,rosi2014precision,fattori2003towards,fixler2007atom,quinn2000measuring,bertoldi2006atom,prevedelli2014measuring}. Atom interferometers are used for testing the Einstein's Equivalence Principle \cite{fray2004atomic,rosi2017quantum,will2006living,will2018theory}. Being massive particles with a well known sensitivity to environmental conditions, atoms are excellent probes for testing the Universality of Free Fall \cite{williams2016quantum,chiow2017gravity,tarallo2014test,schlippert2014quantum,zhou2015test,duan2016test}. Space missions have been proposed to study gravitational waves, dark matter and fundamental aspects of gravity using ultracold atoms \cite{tino2019sage,ElNeaj2020}. Accelerometers based on atom interferometry have been employed for many practical applications including metrology, geodesy, geophysics \cite{agost2009}, engineering prospecting and inertial navigation \cite{cronin2009optics,durfee2006long,peters2001high}. From a technological point of view, building new transportable devices \cite{menoret2018gravity,yong1982transportable} implies remarkable efforts in designing compact optical and electronic systems as well as robust laser locks \cite{schmidt2011portable}. Ongoing studies show that the outer space environment will allow to take full advantage of the potential sensitivity of atom interferometers \cite{bresson2006quantum}.
\par
Among the fundamental constants, the Newtonian gravitational constant has the largest measurement uncertainty. If we take into account all the measurements, starting with the historical experiment held in 1798 by Cavendish \cite{cavendish1798xxi}, the uncertainty on $G$ has improved by less than three orders of magnitude in almost 200 years. Before the first high precision measurement of $G$ using atom interferometry \cite{rosi2014precision,prevedelli2014measuring}, most of the experiments \cite{armstrong2003new,luo2009determination,quinn2013improved,newman2014measurement,li2018measurements} were based on a torsion pendulum or a torsion balance setup, in the same vein as the experiment designed by Cavendish. A new experiment for measuring $G$ at 10 ppm using cold atom interferometry has recently been proposed \cite{rosi2017proposed}.
\par
In this work, we study the concept presented in \cite{rosi2017proposed} and numerically simulate the atom interferometry measurement to optimize the source mass design, maximize the gravity gradiometer signal, and minimize the noise on the interferometric phase. Our simulations show that the setup is compatible with a $G$ measurement having a fractional uncertainty at the $10^{-5}$ level.

\section{Experimental Setup}


\begin{figure*}
  \includegraphics[width=1.00\textwidth]{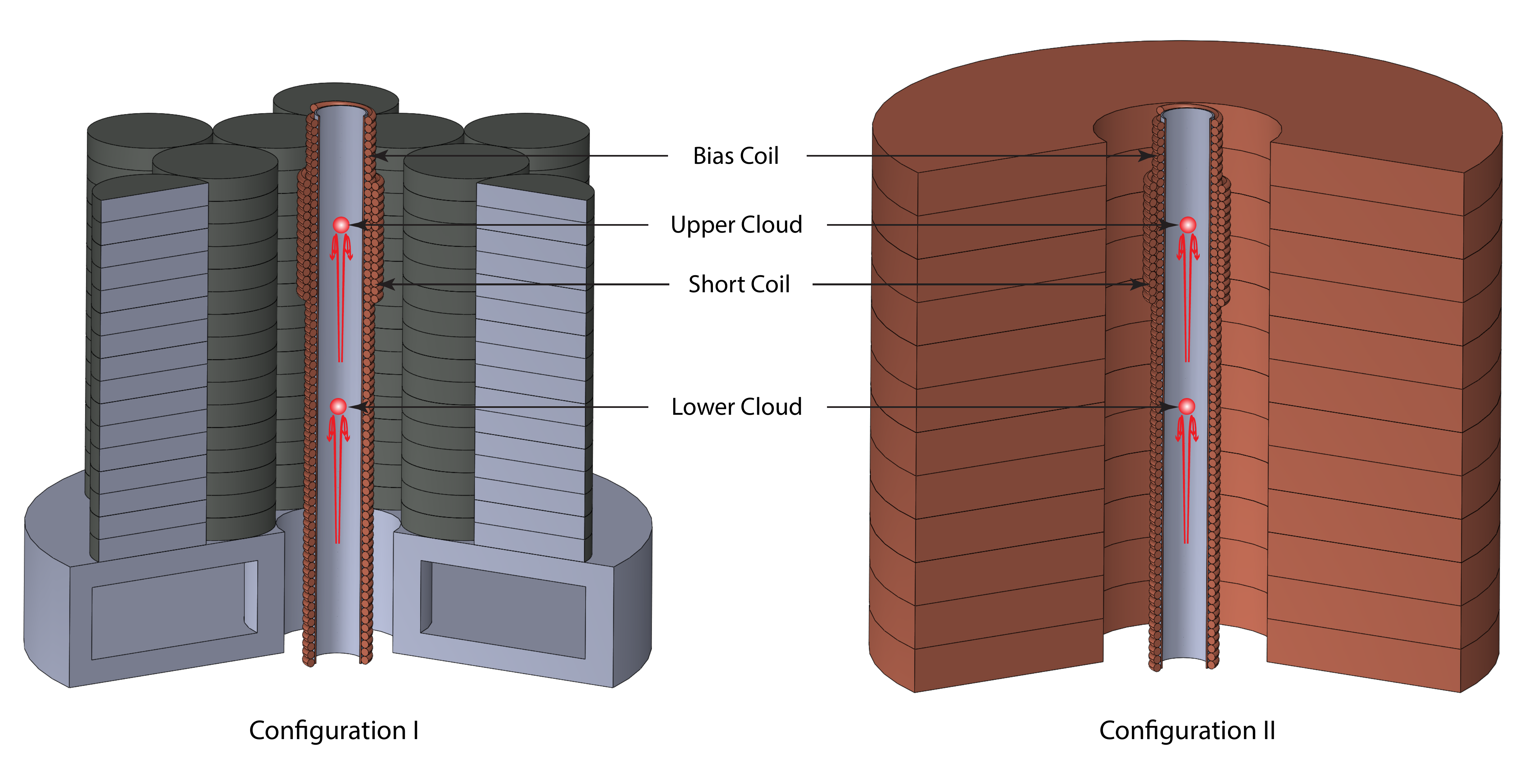}
\caption{Experimental setup showing the vertical tube surrounded by the solenoid and the atomic clouds used to probe the gravitational field generated by the source masses. Two source mass configurations are considered here: (Left) The source mass is composed of 12 tungsten cylinders arranged around the vertical tube in hexagonal geometry. Each cylinder is resulting from 16 tungsten disks stacked together and supported by an aluminium platform. The hollow baseplate assumes a honeycomb structure. (Right) The source mass is a structure composed of 12 copper hollow cylinders stacked together and surrounding the vertical tube.}
\label{fig:1}       
\end{figure*}

In this Section, we describe the experimental setup and the measurement sequence. Fig. \ref{fig:1} shows a detail of the atomic fountain. The cold atom samples used to probe the gravity field are freely falling in the hollow vertical tube of the fountain, which is surrounded by a solenoid defining the bias magnetic field seen by the atoms. Two different source mass configurations symmetrically arranged along the axis of the central vertical tube are studied. They are specifically designed to produce a linear gravitational acceleration profile along the vertical direction.

The experiment uses a Mach-Zehnder gravity gradiometer to measure the gravitational field experienced by the freely falling atoms. Two spatially separated atomic clouds of $^{87}$Rb in free fall along the vertical axis are simultaneously interrogated by counter-propagating Raman lasers \cite{d2019measuring,rosi2014precision,rosi2017quantum}. In our experiment, we probe rubidium atoms on the dipole transition $5S_{1/2} \xrightarrow[]{} 6P_{3/2}$ instead of the usual D2 line. The photon momentum transferred on this transition is taken into account in our numerical simulation when calculating the atomic trajectories during the interferometer sequence.
The atom interference fringes detected at the upper and lower interferometers have a fixed phase difference depending on the gravity gradient experienced by the atoms. Any phase noise induced by vibrations at the reference mirror used to retro-reflect the Raman lasers is in common mode to both interferometers and it can be efficiently rejected. As a result, when plotting the interference fringes detected at the upper interferometer as a function of the fringes at the lower interferometer, measurement points distribute along an ellipse whose eccentricity and rotation angle provide a measurement of the differential phase shift $\Phi$, which is proportional to the average gravity gradient $\Gamma$ along the gradiometer baseline.

As theoretically predicted in \cite{roura2017circumventing} and experimentally demonstrated in \cite{d2017canceling}, a modulation of the effective Raman wavevector during the central $\pi$-pulse of the Mach-Zehnder interferometer can be used to introduce a phase shift which is equivalent to a fictitious gravity gradient $\Gamma^{*}$. In particular, the following relationship holds:
\begin{equation}\label{eq0}
\Gamma^{*} = \frac{-2\Delta k_{\rm eff}}{k_{\rm eff}T^2} = \frac{-8\pi\Delta \nu}{ck_{\rm eff}T^2}
\end{equation}
where $T$ is the free evolution time of the interferometer, $\Delta\nu$ is the frequency detuning of the Raman lasers and $k_{\rm eff}$ represents the Raman effective wavevector.
In this way, a well-controlled phase shift $\Phi^{*}$ can be added to cancel the total gradiometer phase $\Phi_{r}$:
\begin{equation}\label{eq1}
\Phi_{r} = \Phi+\Phi^{*} = k_{\rm eff}T^2 \Bigg(\Gamma+\frac{8\pi}{c}\Delta\nu\Bigg)(d+\Delta v_{0z}T)
\end{equation}
where $d$ and $\Delta v_{0z}$ are the differences in the initial vertical position and velocity of the atomic samples. The frequency detuning $\Delta\nu_{0}$ for which $\Phi_{r}\simeq0$ provides through Eq.~\ref{eq0} a preliminary measurement of the average gravity gradient $\Gamma$ along the atomic gradiometer trajectories. More importantly, the sensitivity of the residual phase $\Phi_r$ to the relative position and velocity of the atomic clouds is significantly reduced as it only depends on the linearity of the gravitational acceleration generated by the source mass and the precision with which the frequency detuning $\Delta\nu_{0}$ cancelling the residual phase $\Phi_r$ can be measured.

We can now describe the experimental strategy foreseen for the $G$ measurement. Following the procedure discussed in \cite{d2017canceling}, we start with a preliminary determination of $\Delta\nu_{0}$ in the presence of the source mass ($\Delta\nu_{0}^{C}$ for the ``close position'') and after moving it far away from the interferometer regions ($\Delta\nu_{0}^{F}$ for the ``far position''). Afterwards, we proceed with the measurement of the residual phases $\Phi_{r}^{F}$ and $\Phi_{r}^{C}$ by periodically alternating the source mass between the close and far position to efficiently reject the local gravity field and any slowly varying phase drifts. The Newton’s constant of gravity $G$ is finally extracted by comparing the measured differential phase $\Phi_{r}^{FC}=\Phi_{r}^{F}-\Phi_{r}^{C}$ with the result of the Monte Carlo simulation that models the experiment and the source mass geometry, and accounts for the measured $\Delta\nu^{FC}_{0}=\Delta\nu_{0}^{F}-\Delta\nu_{0}^{C}$.

It is worth recalling here that when the gradiometer phase shift approaches zero, the least-squares fitting procedure fails in accurately retrieving the ellipse phase angle in the presence of noisy data. To avoid this problem, an external and well-controlled phase shift is added to bring the gradiometer phase close to $\pi/2$ and open the ellipse. To this purpose, a localized and uniform magnetic field pulse is applied during the second part of the interferometer of the upper atomic sample only. Even if this additional phase is a common-mode effect not depending on the source mass position, a careful design of the coils generating the magnetic field pulse and the current generator is important to minimize the measurement noise.

\section{Source mass configuration}
The design of the source masses plays a crucial role in the $G$ measurement. Important aspects to be taken into account are: the high density of the material to produce a large measurement signal; its mechanical properties to ensure a good machinability and a precise finishing of the parts; the geometry (spheres, cylinders, etc.), important to simplify the manufacturing process and to ensure precise modelling of the gravitational field; a regular spatial arrangement to ease the positioning; a high density homogeneity for an accurate modelling of the mass distribution and control of the measurement systematics.

Following the experience gained with the setup used for our previous $G$ measurement~\cite{rosi2014precision}, we study source masses with cylindrical symmetry (disks and hollow cylinders) based on two different materials, tungsten alloy and pure copper. Two different geometries are presented and pros and cons are discussed for each of them.

\subsection{Configuration I: tungsten alloy disks}
The main advantage of tungsten alloy is its very high density ($\rho\approx18300$ kg/m$^3$). This ensures a large differential phase $\Phi$ for a small gravity gradiometer baseline and a short free evolution time. Unfortunately, tungsten alloys have a major drawback, i.e. the poor density homogeneity (typically $10^{-4}-10^{-3}$) due to the thermal gradients that inevitably are still present after the sintering process.
Since the density homogeneity of tungsten alloy parts degrades as their size increases, the source mass is composed of several small units of tungsten that are assembled together.
As shown in Fig. \ref{fig:1}(Left), 16 disks (10~cm radius and 3.75~cm height) are stacked together to form a cylinder with a height of 60~cm; 12 of these cylinders, resulting in a total mass of $4$~t, are precisely positioned around the vertical tube with hexagonal symmetry~\cite{lamporesi2007source}. The overall mass distribution defines a region of $\sim$0.5~m of length along the symmetry axis with a gravity gradient homogeneous better than 2\%. The disk height-to-radius ratio has been tuned to maximize the linearity of the acceleration profile. The reduced mass of each element ($\approx21$~kg) allows to easily move and precisely position the parts. They can be rotated around their own symmetry axis and can be interchanged with each other, important to study the systematic effects due to density inhomogeneities in the bulk material.
Finally, a large platform is used to hold the weight of the 192 tungsten disks. Unfortunately, this element breaks the symmetry of the source mass and perturbs the gravitational field experienced by the atoms. A lightweight ($\approx220$ kg) aluminium honeycomb baseplate has been designed to minimize this effect and reduce the deformations ($>10$ $\mu$m) of the structure under the heavy weight of the source mass. For the purpose of the numerical simulations, the platform has been modeled as a hollow cylinder (0.1~m internal radius, 0.5~m external radius, and 0.2~m height) with a concentric cavity (0.15~m internal radius, 0.45~m external radius, and 0.12~m height) that hosts the honeycomb structure.


\subsection{Configuration II: copper hollow cylinders}
Copper presents several interesting features as a source mass material for a $G$ measurement: it can be produced in large sizes with a high degree of purity and homogeneity; it can be hard-tempered to improve its machinability; it is a quasi-noble metal with a high resistance against corrosion and oxidation; among easily machinable metals, it has a discrete density ($\rho\approx8960$ kg/m$^3$), compatible with the experiment proposed here.
While in our previous measurement of $G$~\cite{rosi2014precision} a high-density source mass material was crucial to cancel the Earth’s gravity gradient around the apogees of the atomic trajectories, this is no longer required in the new experimental scheme thanks to excellent control on gravity gradients that can be achieved by detuning the Raman lasers during the atom interferometry sequence.
However, due to the much lower density with respect to tungsten, the overall size of the copper source mass needs to be re-scaled to maintain the same differential phase of the gravity gradiometer signal. Since $\Phi$ is proportional to the distance between the atomic samples, a reduction of the average gravity gradient $\Gamma$ along the vertical axis by a factor $\eta$ can be compensated by increasing the gradiometer baseline, the interferometer free evolution time, the height of the copper source mass, and consequently its total mass by $\sqrt{\eta}$. Having this in mind, Fig. \ref{fig:1}(Right) shows the second source mass configuration, composed of 12 copper hollow cylinders (0.15~m internal radius, 0.52~m external radius, 0.07~m height) for a total mass of about $6$~t, stacked around the symmetry axis of the atomic fountain tube. Also in this case, the linearity of the acceleration profile generated by the source mass has been optimized by finely tuning the ratio of the internal-to-external radius of the hollow cylinders.
The large copper elements ($\approx500$~kg) do not require the wide and rigid aluminium platform holding the tungsten source masses (stringent requirement in configuration I), which remains difficult to be modelled for the $G$ measurement. On the other hand, the procedure to install and position them is more complex due to the weight and size of the copper elements. Moreover, due to the increased interferometer time, the expansion time of the atomic clouds is significantly longer with respect to the tungsten source mass configuration, thus leading to a degraded signal-to-noise ratio at detection.


\section{Numerical simulations}

In this Section, we describe how the experiment is modelled, the methods of our Monte Carlo simulation, and the parameters that are analysed to optimize the setup. Finally, the measurement sensitivity to the main systematic effects, which depend on the atomic cloud parameters (size and temperature) and initial conditions (position and velocity), is characterized.

\subsection{Source mass gravity field}
Our analysis is not considering the Earth's gravity field contribution as it is efficiently cancelled when taking differential measurements with the source mass alternatively in the ``close'' and ``far'' position. Consequently, $\Phi_{r}\simeq\Phi_{r}^{FC}$ and $\Delta\nu_{0}\simeq\Delta\nu^{FC}_{0}$.

The gradiometer phase $\Phi$ generated by the source mass distribution is calculated with the perturbative treatment already presented in previous works \cite{prevedelli2014measuring}. The gravitational potential $U_{1}$ generated by source mass can be considered a small perturbation of the large background potential $U_{0}$ of the Earth. Therefore,
\begin{equation}
\Phi = \frac{m_{Rb}}{\hbar}\Bigg(\oint_{up}U_1(t)dt-\oint_{dw}U_1(t)dt\Bigg)
\end{equation}
where the integrals are evaluated on the closed paths defined by the wave packet trajectories of the upper and lower interferometers due to the $U_{1}$ potential only. The calculation of $U_{1}$ is optimized based on the shape and the dimensions of the elements composing the source mass. For the disks in configuration I, a multipole expansion \cite{prevedelli2014measuring} using the first three terms is sufficient to reach the required accuracy ($<$ 1 $\mu$rad with respect to the exact formula) thanks to the small height-to-radius ratio. For the large hollow cylinders of configuration I (parts from the aluminium platform) and configuration II (copper elements), the exact formula can be approximated by a second-order expansion in the radial coordinate \cite{prevedelli2014measuring} considering that the atoms are within a few millimeters from the symmetry axis.

Finally, $\Phi_{r}$ is computed after implementing the gravity gradient compensation procedure (see Eq.~\ref{eq1}). $\Delta\nu_{0}$ can be precisely evaluated by a single particle simulation along the vertical axis using the nominal coordinates of the cloud barycentres. However, as this is an important experimental parameter that is known with a limited accuracy, we also vary $\Delta\nu_{0}$ by adding a relative offset which corresponds to 0.1\%, 0.5\%, and 1.0\% of the calculated value.

\subsection{Atomic gradiometer configuration}
\label{sub1}
The atomic gradiometer configuration is defined by the following single-particle parameters: gradiometer baseline $d$, interferometer free evolution time $T$, atomic positions and velocities at the first beam splitter pulse. The first two parameters determine the instrument sensitivity and therefore need to be maximized. If $L$ is the length of the spatial region where the vertical gravity gradient $\Gamma$ is homogeneous ($<$ 2\% over a 0.5~m distance), $d\approx0.5gT^2\approx L/2$ defines the optimal condition, where $g$ is the local gravity acceleration. The initial vertical velocity $v_{0z}$ of the lower and upper clouds at the start of the interferometric sequence is $g(T+\Delta t)$, where $\Delta t\approx1$~ms marks the difference between the instant in which the atomic clouds reach the apogees of their trajectories and the time of the central $\pi$ pulse of the interferometer. This time difference is needed to drive a velocity selective Raman transition around the apogees, thus avoiding degeneracy with the transition having opposite $k_{\rm eff}$. Finally, the vertical positions $z_0$ and $z_0+d$ of the upper and the lower atomic clouds at the start of the interferometer (first $\pi/2$ pulse) can be determined by imposing
\begin{equation}
\frac{\partial\Phi}{\partial z_0} = -\frac{m_{Rb}}{\hbar}\Bigg(\oint_{up}a(t)dt-\oint_{dw}a(t)dt\Bigg) = 0
\end{equation}
where $a(t)$ is the gravitational acceleration induced by the source mass. In this way, phase fluctuations due to the vertical position of the atomic clouds can be canceled to first order. Table \ref{tab:1} reports the optimized parameters for the two source mass configurations.

Then, a Monte Carlo simulation of the experiment evaluates both the gradiometer phase $\Phi_r$ and its error by varying the initial atom position and velocity according to the density and velocity distribution of the two atomic clouds.
In this way, it is possible to both optimize the $G$ measurement with respect to the experimental parameters and to study how each parameter individually contributes to the overall error budget. Tables \ref{tab:2} and \ref{tab:3} list the key parameters and the values used in the numerical simulation for the two configurations of the source masses. Those figures are typical for thermal clouds produced in an atomic fountain after velocity selection \cite{rosi2014precision,prevedelli2014measuring}.

\begin{table}
\caption{ Optimized single-particle parameters for the one-dimensional atomic gradiometer along the z-axis. The z-coordinate origin corresponds to the position of the source mass barycenter.}
\label{tab:1}       
\begin{tabular}{lll}
\hline\noalign{\smallskip}
Parameter(Units) & Configuration I & Configuration II  \\
\noalign{\smallskip}\hline\noalign{\smallskip}
$z_{0}$(mm) & -161.7 & -232.2 \\
$T$(ms) & 220 & 263 \\
$d$(mm) & 230 & 329 \\
$v_{0}$(m/s) & 2.167 & 2.593 \\
\noalign{\smallskip}\hline
\end{tabular}
\end{table}

\begin{table}
\caption{ The atomic cloud parameters and the corresponding values explored in the Monte Carlo simulations for configuration I. The z-coordinate origin is at the source mass barycenter. Velocity and position are described by Gaussian distributions with radial symmetry. The first seven parameters define the position and the velocity of the atomic cloud barycenter, the remaining four describe the expansion dynamics of the sample. Note that $z_{u} = z_{0} + d$.}
\label{tab:2}       
\begin{tabular}{ll}
\hline\noalign{\smallskip}
Parameters & Variation \\
\noalign{\smallskip}\hline\noalign{\smallskip}
$x$ (mm) & $-1.0, -0.5, 0.0, 0.5, 1.0$ \\
$y$ (mm) & $-1.0, -0.5, 0.0, 0.5, 1.0$ \\
$z_{dw}$ (mm) & $z_0-0.2, z_0-0.1, z_0, z_0+0.1, z_0+0.2$ \\
$z_{up}$ (mm) & $z_u-0.2, z_u-0.1, z_u, z_u+0.1, z_u+0.2$ \\
$v_{x}$ (mm/s) & $-1.0, -0.5, 0.0, 0.5, 1.0$ \\
$v_{y}$ (mm/s) & $-1.0, -0.5, 0.0, 0.5, 1.0$ \\
$v_{z}$ (mm/s) & $v_{0}-0.2, v_{0}-0.1, v_{0}, v_{0}+0.1, v_{0}+0.2$ \\
$\sigma_{r}$ (mm) & $2.6, 2.8, 3.0, 3.2, 3.4$  \\
$\sigma_{z}$ (mm) & $2.8, 2.9, 3.0, 3.1, 3.2$ \\
$\sigma_{v_{r}}$ (mm/s) & $14, 17, 20, 23, 26$ \\
$\sigma_{v_{z}}$ (mm/s) & $2.8, 2.9, 3.0, 3.1, 3.2$ \\
\noalign{\smallskip}\hline
\end{tabular}
\end{table}

\begin{table}
\caption{ The atomic cloud parameters and the corresponding values explored in the Monte Carlo simulations for configuration II. The z-coordinate origin is at the source mass barycenter. Velocity and position are described by Gaussian distributions with radial symmetry. The first five parameters define the position and the velocity of the atomic cloud barycenter, the remaining four describe the expansion dynamics of the sample. Note that $z_{u} = z_{0} + d$. The number of parameters is reduced with respect to configuration I due to the cylindrical symmetry .}
\label{tab:3}       
\begin{tabular}{ll}
\hline\noalign{\smallskip}
Parameters & Variation \\
\noalign{\smallskip}\hline\noalign{\smallskip}
$r$ (mm) & $0.0, 0.5, 1.0, 1.5, 2.0$ \\
 $z_{dw}$ (mm) & $z_0-0.2, z_0-0.1, z_0, z_0+0.1, z_0+0.2$ \\
 $z_{up}$ (mm) & $z_u-0.2, z_u-0.1, z_u, z_u+0.1, z_u+0.2$ \\
 $v_{r}$ (mm/s) & $0.0, 0.5, 1.0, 1.5, 2.0 $ \\
 $v_{z}$ (mm/s) & $v_{0}-0.2, v_{0}-0.1, v_{0}, v_{0}+0.1, v_{0}+0.2$ \\
 $\sigma_{r}$ (mm) & $2.6, 2.8, 3.0, 3.2, 3.4$ \\
 $\sigma_{z}$ (mm) & $2.8, 2.9, 3.0, 3.1, 3.2$ \\
 $\sigma_{v_{r}}$ (mm/s) & $14, 17, 20, 23, 26$ \\
 $\sigma_{v_{z}}$ (mm/s) & $2.8, 2.9, 3.0, 3.1, 3.2$ \\
\noalign{\smallskip}\hline
\end{tabular}
\end{table}

\subsection{Magnetic fields}
As already explained earlier, we need to add an external well-controlled phase shift $\Phi_B\approx\pi/2$ to open the ellipse and correctly retrieve the gradiometer phase. This extra phase is induced by a magnetic field pulse (typically a few ms duration) acting on one interferometer only, after the central $\pi$ pulse. Due to the second-order Zeeman effect,
\begin{equation}
\Phi_{B} = 2\pi\alpha \Bigg(\oint_{up}B(t)^2dt-\oint_{dw}B(t)^2dt\Bigg)
\end{equation}
where $\alpha = 575.15$ Hz$/G^{2}$. The magnetic field experienced by the atoms
\begin{equation}
B(t) = B_{bias}(t)+[\theta(t-T)-\theta(t-T-\Delta t)]B_{short}(t)
\end{equation}
where $\theta(t)$ is the Heaviside step function, is the combination of the bias magnetic field $B_{bias}(t)$ produced by the solenoid surrounding the vertical tube and the magnetic field pulse $B_{short}(t)$ produced by an additional short magnetic coil acting on the upper interferometer (see Fig. \ref{fig:1}). This coil is only powered for a short duration $\Delta t$ after the central $\pi$ pulse.

For the $G$ measurement, it is important to control any instabilities that might be introduced by $\Phi_{B}$ in the gravity gradiometer phase. In particular, the $\Phi_{B}$ sensitivity to the position of the atomic samples shall be minimised. For this purpose, we optimize the length and the position of the short coil. Optimal parameters are extracted from the short gradiometer baseline of configuration I. Indeed, the same parameters will reduce even further the $\Phi_{B}$ sensitivity to atoms position when used in configuration II due to the longer gradiometer baseline. To determine the short coil length $l$, we run a single particle simulation along the z-axis with the coil center coinciding with the apogee of the upper atomic cloud and vary $l$ to minimise the expression
\begin{equation}
\begin{split}
&\chi(l) = \\
&\sqrt{\Bigg(\left.\frac{\partial\Phi_B(z_{up},l)}{\partial z_{up}}\right|_{z_{up}=z_0}\Bigg)^2 + \left.\Bigg(\frac{\partial\Phi_B(z_{dw},l)}{\partial z_{dw}}\right|_{z_{dw}=z_0+d}\Bigg)^2}
\end{split}
\end{equation}
with the constrain $\Phi_B\approx\pi/2$.
Finally, the short coil position has been optimized by searching for an extremum of $\Phi_B$:
\begin{equation}\label{eq2}
 \frac{\partial\Phi_B}{\partial z_B} = 0
\end{equation}
where $z_B$ defines the vertical coordinate of the short coil center. All relevant parameters are reported in Table \ref{tab:4}.
Once the parameters of the short coil are defined, the overall magnetic field is calculated by using the analytical expressions for a finite solenoid in 3D.

\begin{table}
\caption{ Optimized parameters for generating the magnetic field phase $\Phi_B$. The z-coordinate origin is set in correspondence of the bias coil center.}
\label{tab:4}       
\begin{tabular}{ll}
\hline\noalign{\smallskip}
Parameters & Value \\
\noalign{\smallskip}\hline\noalign{\smallskip}
Bias coil length & 4 m \\
Short coil length & 0.230 m \\
Number of turns/meter & 1000 \\
Bias coil current & 9.9 mA  \\
Short coil current & 19 mA \\
$\Delta$$t$ & 4 ms \\
$z_B$ & -1.373 m \\
$z_0$ & -1.850 m  \\
\noalign{\smallskip}\hline
\end{tabular}
\end{table}

\section{Results}

The Monte Carlo simulations of the experiment were run by varying one parameter at a time and keeping the others fixed at their central values. We launch 1000 atoms per cloud, which ensures a resolution of 2 - 3~$\mu$rad in the residual phase $\Phi_r$. This is sufficient to reach the 40~$\mu$rad level in the optimization of the magnetic field phase $\Phi_B$. The maximum-to-minimum phase differences in $G$ units are reported in Tables \ref{tab:5} to \ref{tab:7}.

\begin{table}
\caption{ Maximum-to-minimum phase difference of $\Phi_r$ expressed in $G$ units for the source mass in configuration I and for different levels of gravity gradient compensation. The simulation standard error is equal to 3~$\mu$rad, equivalent to $2\times 10^{-6}~G$.}
\label{tab:5}       
\begin{tabular}{llll}
\hline\noalign{\smallskip}
& Max-to-min &  Max-to-min &  Max-to-min \\
Variables  & diff.  &  diff. & diff. \\
& 0.1\% comp.  &  0.5\% comp. & 1\% comp. \\
 &($\times 10^{-6}G$) & ($\times 10^{-6}G$) & ($\times 10^{-6}G$) \\
\noalign{\smallskip}\hline\noalign{\smallskip}
$x$ & 5 & 5 & 5 \\
$y$ & 4 & 4 & 4 \\
$z_{dw}$ & 6 & 7 & 8 \\
$z_{up}$ & 21 & 22 & 23 \\
$v_{x}$ & 5 & 5 & 5 \\
$v_{y}$ & 3 & 3 & 3 \\
$v_{z}$ & 4 & 4 & 4\\
$\sigma_{r}$ & 16 & 17 & 18 \\
$\sigma_{z}$ & 12 & 12 & 12 \\
$\sigma_{v_{r}}$ & 59 & 59 & 59 \\
$\sigma_{v_{z}}$ & 8 & 8 & 8 \\
\noalign{\smallskip}\hline
\end{tabular}
\end{table}

\begin{table}
\caption{ Maximum-to-minimum phase difference of $\Phi_r$ expressed in $G$ units for the source mass in configuration II and for different levels of gravity gradient compensation. The simulation standard error is equal to 1.5~$\mu$rad, equivalent to $1\times 10^{-6}~G$.}
\label{tab:6}       
\begin{tabular}{llll}
\hline\noalign{\smallskip}
& Max-to-min &  Max-to-min &  Max-to-min \\
Variables  & diff.  &  diff. & diff. \\
& 0.1\% comp.  &  0.5\% comp. & 1\% comp. \\
 &($\times 10^{-6}G$) & ($\times 10^{-6}G$) & ($\times 10^{-6}G$) \\
\noalign{\smallskip}\hline\noalign{\smallskip}
 $r$ & 3 & 3 & 3 \\
 $z_{dw}$ & 6 & 7 & 8 \\
 $z_{up}$ & 5 & 6 & 7 \\
 $v_{r}$ & 1.4 & 1.6 & 1.8 \\
 $v_{z}$ &1.0 & 1.1 & 1.3  \\
 $\sigma_{r}$ & 3 & 3 & 3 \\
 $\sigma_{z}$ & 3 & 3 & 3 \\
 $\sigma_{v_{r}}$ & 20 & 20 & 20 \\
 $\sigma_{v_{z}}$ & 1.3 & 1.7 & 2 \\
\noalign{\smallskip}\hline
\end{tabular}
\end{table}

\begin{table}
\caption{Maximum-to-minimum phase differences of $\Phi_B$ expressed in $G$ units for configuration I. Configuration II shows smaller peak-to-peak values due to the longer gradiometer baseline. The standard error is equal to 45$~\mu$rad, equivalent to $31\times 10^{-6}G$.}
\label{tab:7}       
\begin{tabular}{ll}
\hline\noalign{\smallskip}
Variables & Max-to-min \\
& diff. \\
& ($\times 10^{-6}G$) \\
\noalign{\smallskip}\hline\noalign{\smallskip}
 $r$ & 36\\
 $z_{dw}$ & 69 \\
 $z_{up}$ & 65 \\
 $v_{r}$ & 57 \\
 $v_{z}$ & 99  \\
 $\sigma_{r}$ & 36 \\
 $\sigma_{z}$ & 111\\
 $\sigma_{v_{r}}$ & 63 \\
 $\sigma_{v_{z}}$ & 21 \\
\noalign{\smallskip}\hline
\end{tabular}
\end{table}

$\Phi_B$ shows no clear trend with respect to the simulation parameters. The maximum variations are always within $10^{-4}~G$ for both configuration I and II. Based on our previous experience, the typical stability of an atomic fountain setup and a careful design of the magnetic field coils allow to easily integrate $\Phi_B$ down to $10^{-4}$ both in the close and far position of the source mass.
$\Phi_r$ does not depend on the level of gravity gradient compensation in both configurations. Therefore, a preliminary determination of the gravity gradient within 1\% is sufficient for the experiment. Simulations have not revealed any clear dependence of $\Phi_r$ as a function of the experimental parameters, with the exception of the width of the radial velocity distribution $\sigma_{v_{r}}$. In this case, we find linear slopes of 7~$\mu$rad/(mm~s$^{-1}$) and 2.6~$\mu$rad/(mm~s$^{-1}$) for configuration I and II, respectively. Assuming a 3~mm/s error on $\sigma_{v_{r}}$~\cite{prevedelli2014measuring}, we obtain relative uncertainties on $G$ of $13\times 10^{-6}$ and $5\times 10^{-6}$ for configurations I and II.

As expected, configuration II provides the best result. Indeed, as discussed before, the absence of a platform holding the source masses improves the homogeneity of the gravity gradient along the vertical axis; moreover, the larger size and the lower density of the copper source mass reduce the radial gravity gradient thus providing a better control on the systematic effects that depend on the radial position of the atomic samples .

\section{Conclusions}
In this paper, we have studied a new method to measure the Newtonian gravitational constant by atom interferometry.

Monte Carlo simulations are used to optimize the source mass configuration and to estimate the contribution of critical parameters such as position and velocity of the atomic clouds as well as their temperature and size to the $G$ measurement. As shown in previous work \cite{rosi2014precision}, they represent the most important source of systematic error, currently preventing from reaching the $10^{-5}$ uncertainty level in atom interferometry measurements of $G$.

Our study compares two different source mass designs, the first composed of tungsten alloy disks symmetrically arranged on a moving platform, the second using large copper elements stacked together to form a hollow cylinder. The copper source mass clearly provides the best result, thanks to: the higher density homogeneity of copper with respect to sintered tungsten; the larger source mass size and the lower density of the copper that reduce the measurement sensitivity to radial displacements of the atomic clouds; the higher homogeneity of the gravity gradient along the vertical axis that could be obtained thanks to the absence of an aluminium support platform holding the source masses. 

The simulation results show that our measurement concept, relying on the compensation of the gravity gradient effects \cite{d2017canceling}, is compatible with a $G$ measurement at the $10^{-5}$ uncertainty level.

\section*{Acknowledgments}
We acknowledge financial support from INFN and the Italian Ministry of Education, University and Research (MIUR) under the aegis of Progetto Premiale ``Interferometro Atomico'' and PRIN 2015. GR acknowledges financial support from the European Research Council, Grant No. 804815 (MEGANTE).

\section*{Authors contributions}

All the authors were involved in the preparation of the manuscript. All the authors have read and approved the final manuscript.

%
\bibliographystyle{epj}
\bibliography{aipsamp}
%
%
%

\end{document}